\documentclass[fleqn,usenatbib]{mnras}
\usepackage{amsmath}
\usepackage{graphicx}
\usepackage{amssymb,txfonts}
\usepackage{url}

\newcommand{\msun}{{\rm M}_{\sun}}
\newcommand{\rsun}{{\rm R}_{\sun}}
\newcommand{\lsun}{{\rm L}_{\sun}}
\newcommand{\source}{GRS 1915+105}

\newbox\grsign \setbox\grsign=\hbox{$>$} \newdimen\grdimen \grdimen=\ht\grsign
\newbox\simpropbox
\setbox\simpropbox=\hbox{\raise.5ex\hbox{$\propto$}\llap
     {\lower.5ex\hbox{$\sim$}}}\ht2=\grdimen\dp2=0pt
\def\simprop{\mathrel{\copy\simpropbox}}

\title[V1487 Aql/GRS 1915+105]{The mass, luminosity and mass-loss rate of the donor of\\ the V1487 Aql/GRS 1915+105 binary system}
\author[J. Zi\'o{\l}kowski and A. A. Zdziarski]{Janusz Zi\'o{\l}kowski\thanks{E-mail:
jz@camk.edu.pl, aaz@camk.edu.pl} and Andrzej A. Zdziarski$^\star$\\
Nicolaus Copernicus Astronomical Center, Polish Academy of Sciences, Bartycka 18, PL-00-716 Warszawa, Poland}
\begin{document}

\date{Accepted 2017 May 2. Received 2017 May 2; in original form 2017 February 25}

\pagerange{\pageref{firstpage}--\pageref{lastpage}} \pubyear{2017}

\maketitle

\label{firstpage}

\begin{abstract}
The donor in the microquasar GRS 1915+105 is a low-mass giant. Such a star consists of a degenerate helium core and a hydrogen-rich envelope. Both components are separated by a hydrogen burning shell. The structure of such an object is relatively simple and easy to model. Making use of the observational constraints on the luminosity and the radius of the donor, we constrain the mass of this star with evolutionary models. We find a very good agreement between the constraints from those models and from the observed rotational broadening and the NIR magnitude. Combining the constraints, we find solutions with stripped giants of the mass of $\geq\!0.28{\rm M}_{\sun}$ and of the spectral class K5 III, independent of the distance to the system, and a distance-dependent upper limit, $\lesssim\!1{\rm M}_{\sun}$. We also calculate the average mass transfer rate and the duty cycle of the system as a function of the donor mass. This rate is much below the critical rate (at which the system would become persistent), and the duty cycle is less than 20 per cent.
\end{abstract}

\section{Introduction}

\source\ is a low mass X-ray binary, which appears to be the most distinct Galactic microquasar \citep{mr94}. Its optical component got a variable-star name V1487 Aql. Its outburst began on 1992 August 15 \citep*{cbl92}, and it has remained in the outburst state since then. The system contains a black hole and a low mass K--M III giant donor \citep{greiner01b,hg04,steeghs13}, and it has a long period of $P=33.85\pm 0.16$ d \citep{steeghs13}. Its donor fills its Roche lobe and supplies the matter accreted by the black hole. The mass of the black hole, $M_1$, can be constrained from the radial velocity amplitude, which was measured by VLT as $K_2= 126\pm 1$ km s$^{-1}$ \citep{steeghs13}. The value of the mass depends on the inclination, $i$, as $M_1\propto (K_2/\sin i)^3$. In the case of \source, the value of $i$ can be determined for the jets only, where it depends on the distance to the source \citep{mr94,fender99}. The distance is currently determined as $d= 8.6^{+2.0}_{-1.6}$ kpc from radio parallax measurements \citep{reid14}, which determination is consistent with an independent estimate of $d\lesssim 10$ kpc by \citet{zdz14} based on considering the jet kinetic power, and which yields $M_1=12.4^{+2.0}_{-1.8}\msun$ \citep{reid14}.

While the mass of the black hole may be considered relatively precisely known, the mass of the K--M III donor is more poorly constrained. Its mass can be constrained by rotational broadening, which has been measured by \citet{steeghs13} as $v\sin i=21\pm 4$ km s$^{-1}$, where the standard deviation $v$ for an individual measurement is $\sim$2--3 km s$^{-1}$, and the total uncertainty includes estimated systematic errors. This value combined with their $K_2$ measurement yields $q\simeq 0.042\pm 0.024$. For the distance-dependent black-hole mass estimate of \citet{reid14}, this gives $M_2\simeq 0.52\pm 0.31\msun$. Thus, the donor has a mass substantially lower than that of an isolated star of the same spectral class \citep{cox00}, i.e., it is a 'stripped giant' (e.g., \citealt*{webbink83}, hereafter WRS83). 

We explore here another possibility to constrain the donor mass based on modelling of its internal structure. Since the donor is not a main sequence star but rather an evolved low-mass giant, its structure is relatively simple and easy to model. Making use of the observational constraints on the luminosity and the radius of the donor, we construct an evolutionary model of this star and attempt to constrain its mass. For that, we also need the effective temperature of the donor. The currently most accurate NIR observations of this system are those of \citet{steeghs13}, who matched their spectra to those of K0, K1, K2, K5 III and M0 III template stars. Thus, we adopt here the possible range of the spectral classes\footnote{\citet{fm15} gave the range of the spectral class of the donor as K0--3 III, but that choice was not based on any additional constraints with respect to those of \citet{steeghs13} (J. McClintock, private communication.) Also, they gave $M_2\simeq 0.58\pm 0.33\msun$, which appears to be due to a typo.} from K0 III to M0 III. 

\section{Observationally determined parameters}
\label{radius}

The third Kepler law and the relation between the donor radius (which is equal to the radius of its Roche lobe) and the separation, $a$, for $M_2\ll M_1$ \citep{bp67} are
\begin{equation}
a =\frac{[G(M_1+M_2)]^{1/3} P^{2/3}}{(2\pi)^{4/3}},\quad R_2=\frac{2 a}{3^{4/3}} \left(\frac{M_2}{M_1+M_2} \right)^{1/3},
\label{kepler}
\end{equation}
respectively. This yields the standard formula,
\begin{equation}
R_2=(2 G M_2)^{1/3}\left(\frac{P}{9\pi}\right)^{2/3} \simeq 1.945\rsun (M_2/\msun)^{1/3} (P/1\,{\rm d})^{2/3}.
\label{r_k}
\end{equation}
This formula makes the radius of the donor filling its Roche lobe one of the most accurately determined parameters of a binary. The orbital period is usually known with high precision and the dependence on $M_2$ is weak. Moreover, equation (\ref{r_k}) does not depend on the distance to the binary system. For discussion of the accuracy of equation (\ref{r_k}) see, e.g., \citet{zdz16}. Given the small uncertainty on the period of \source, its contribution to the error is negligible. 

We note that the Roche-lobe radius for a given donor mass of equation (\ref{r_k}), $R_2(M_2,P)$, is identical to that implied by the rotational broadening,
\begin{equation}
R_2=\frac{P (v\sin i)}{2\upi \sin i},
\label{rad_rot}
\end{equation}
for $M_2=q(v\sin i, K_2) M_1(K_2, P, i, q)$ substituted in the former. Here $q$ follows from the standard rotational-broadening relationship (e.g., \citealt{wh88}), which solution can be found as
\begin{align}
&q=\frac{[r(y)-1]^2}{3 r(y)},\quad r(y)=2^{-1/3}\left[2+27y+3^{3/2}\sqrt{y(4+27y)}\right]^{1/3}\geq 1,\nonumber \\
&y=\left(\frac{3^{4/3} v\sin i}{2 K_2}\right)^3\!,\label{qratio}
\end{align}
while the black-hole mass is
\begin{equation}
M_1=\frac{P K_2^3(1+q)^2}{2\upi G\sin^3 i}.
\label{mbh}
\end{equation}
Equations (\ref{rad_rot}--\ref{qratio}) assume corotation, which is very likely in \source.

We can relate the inclination to the distance by assuming the inner jet has the same direction as the binary axis, which yields $i=\arctan\left[2\mu_{\rm a}\mu_{\rm r} d/ (\mu_{\rm a}-\mu_{\rm r})c\right]$, where $\mu_{\rm a}$ and $\mu_{\rm r}$ are, respectively, the angular velocities of the approaching and receding jet. \citet{zdz14} used the weighted average of the inclination resulting from the observations by \citet{mr94} and \citet{fender99}. On the other hand, \citet{reid14} argued that the jet direction changes between the projected distance from the centre of $0.3''$ (observed by \citealt{fender99}) and $1''$ \citep{mr94}, and thus the results of the former are more relevant for estimating the binary plane orientation, resulting in $i=59^{+5}_{-4}\degr$. 

\citet{steeghs13} argued that the alignment is very likely, and this assumption has indeed been universally used in the mass estimates for \source. On the other hand, an also likely and widely accepted model for the low-frequency QPOs/breaks in the power spectra of black-hole binaries is the Lense-Thirring precession of an inner hot part of the accretion flow (e.g., \citealt*{ingram09,ingram11}). This model requires a misalignment between the black-hole spin and binary axes, though the minimum required misalignment appears not to be specified. \source\ does show low-frequency QPOs, which appear very similar to those in other black-hole binaries (e.g., \citealt{yan13}), and the above model can apply to it. Also, the black-hole binary GRO J1655--40 has an accurate determination of the orbital axis inclination of $68.65\pm 1.5\degr$ \citep{beer02}, which is significantly different from the jet axis inclination of $85\pm 2\degr$ \citep{hjellming95}. If the axes are different in \source\ as well, the alignment-based estimates would become inaccurate, and, e.g., the marked difference between the black-hole mass of \source\ of $M_1=12.4^{+2.0}_{-1.8}\msun$ \citep{reid14} and the average for accreting low-mass black-hole binaries of $7.8\pm 1.2\msun$ \citep{ozel10}\footnote{We note here that the above average needs to be updated, e.g., taking into account the revision of the black-hole mass in Nova Muscae from $7.2.\pm 0.7\msun$ to $11.0^{+2.1}_{-1.4}\msun$ \citep{wu16}.} may disappear.

\begin{figure}
\centerline{\includegraphics[width=\columnwidth]{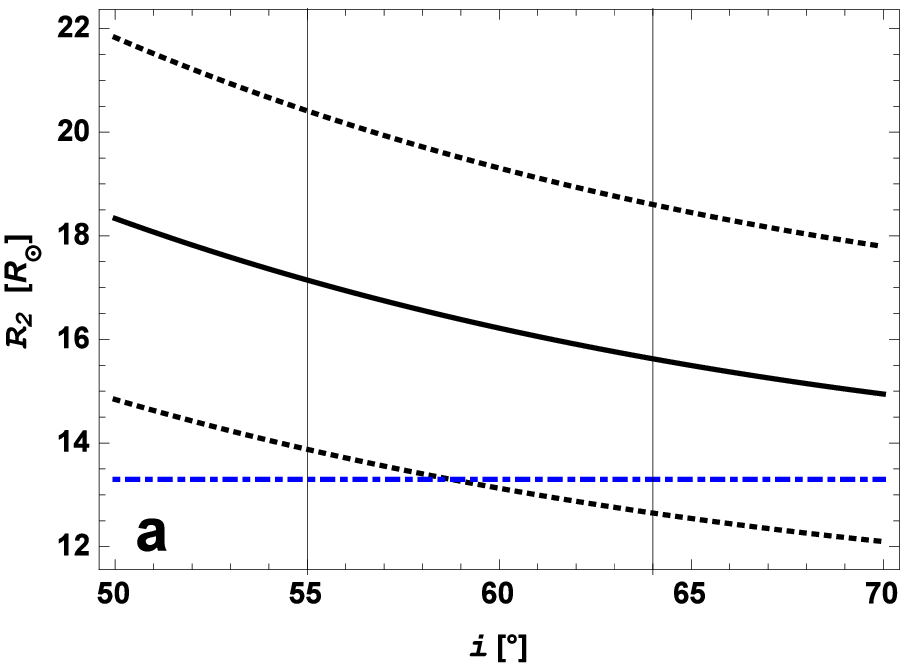}} 
\centerline{\includegraphics[width=\columnwidth]{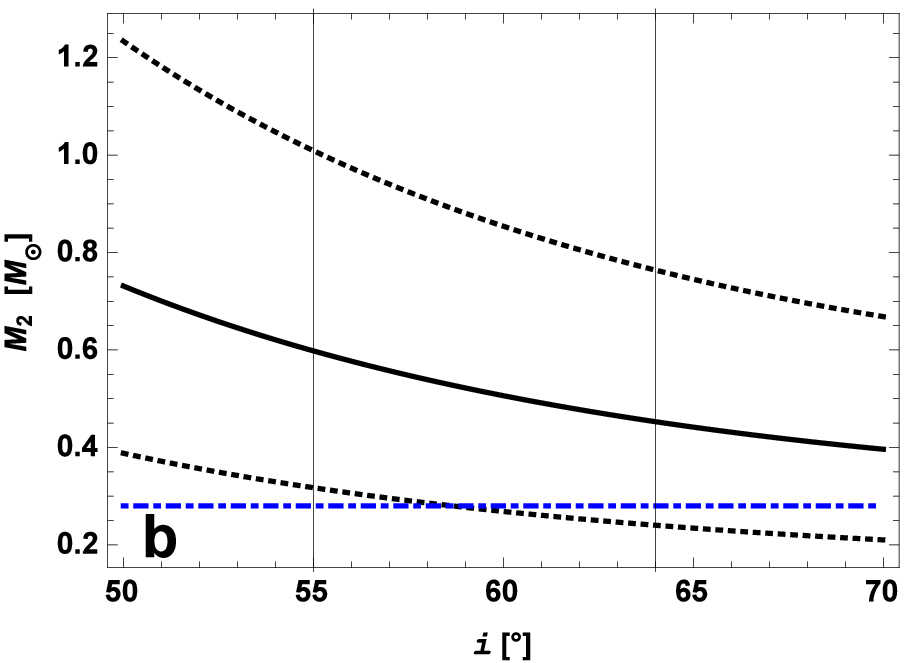}} 
\caption{The donor radius (a) and mass (b) as functions of the binary inclination. The black solid curves give the best-fit values from the observed rotational broadening assuming the Roche-lobe filling and corotation, and the black dotted curves enclose the ranges of uncertainties, which is dominated by the measurement error of $v\sin i$. The range of $i$ compatible with the parallax distance, assuming the alignment of the jet with the binary axis, and using the results of \citet{fender99} is $55\degr$--$64\degr$ \citep{reid14}, as shown by the vertical thin lines. The horizontal blue dot-dashed lines show $M_2=0.28$ and the corresponding radius, which are the minimum possible values found from evolutionary stellar models in Section \ref{models}. The ranges allowed by both the observations and the models are those below the upper dotted curve and above both the dot-dashed and lower dotted curves.
}
\label{r2_i}
\end{figure}

\begin{figure}
\centerline{\includegraphics[width=\columnwidth]{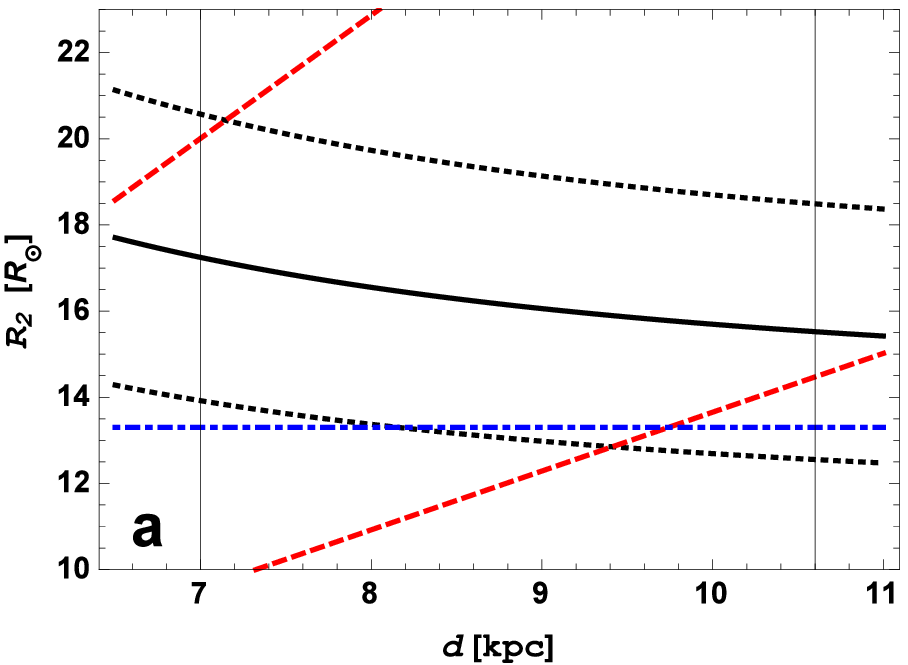}} 
\centerline{\includegraphics[width=\columnwidth]{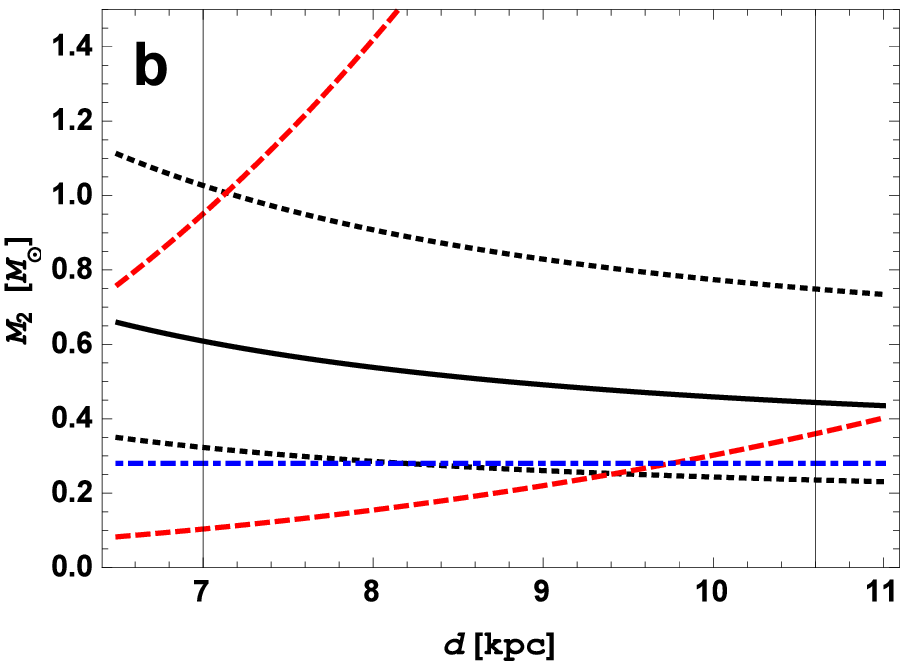}} 
\caption{The donor radius (a) and mass (b) as functions of the distance. The black solid curves show  the best-fit values from the observed rotational broadening assuming the Roche-lobe filling, corotation and alignment of the jet and binary axes, and the black dotted curves enclose the ranges of uncertainties. The range of $d$ from the parallax \citep{reid14} is shown by the vertical thin lines. The red dashed curves enclose the range allowed by the observed NIR flux (approximately independent of $i$). The horizontal blue dot-dashed lines show $M_2=0.28$ and the corresponding radius, which are the minimum possible values based on evolutionary consideration. The allowed parameter region is within the innermost dotted, dashed and dot-dashed curves and the vertical lines.
}
\label{r2_d}
\end{figure}

Given this uncertainty, we show the constraints on the radius and mass of the donor as functions of the inclination directly in Fig.\ \ref{r2_i}, and of the distance (assuming the alignment) in Fig.\ \ref{r2_d}.  In the former, we show the range of the inclinations obtained by \citet{reid14} assuming the alignment. 

We can also constrain the size and mass of the donor vs.\ the distance by using the observed NIR flux of the donor (as first done by \citealt{zdz05}). The unveiled donor K magnitude (at $\lambda=2.2\mu$m) and the extinction towards the system have been estimated as 14.5--15.0 and $2.2\pm 0.3$ by \citet{greiner01b} and \citet{chapuis04}, respectively. This gives the extinction-corrected magnitude of K $\simeq 12.0$--13.1, or the flux per unit wavelength at $2.2\mu$m as (2.2--$6.2)\times 10^{-8}$ erg s$^{-1}$ cm$^{-3}$. By approximating the stellar spectrum as a blackbody at the effective temperature, $B_\lambda (T_{\rm eff})$, i.e., $F_\lambda=\upi B_\lambda(T_{\rm eff}) (R_2/d)^2$, we can then obtain the stellar radius as a function of the distance. The results for the temperatures within the range of $T_{\rm eff}=3690$--4660 K \citep{cox00}, corresponding to the adopted range of the spectral classes of K0 III--M0 III, are shown by the dashed red curves in Fig.\ \ref{r2_d}. These results are basically equivalent, and almost the same as those obtained using the Barnes-Evans relation \citep*{barnes76,beuermann99}, as given for the K magnitude by equation (1a) of \citet{cahn80}, and for the range of the surface brightness of $F_{\rm K}=3.81$--3.86 (which approximately corresponds to K0--M0 giants, see fig.\ 2 of \citealt{cahn80}). 

Combining the above constraints for the allowed range of the distance, we find $12.5\lesssim R_2/\rsun\lesssim 21$ and $0.25\lesssim M_2/\msun\lesssim 1$. The upper limit is larger than that of \citet{steeghs13} because they gave that from propagation of errors, while here we give it for the entire allowed range of $d$. At a given distance, the constraints are more stringent, as shown on Fig.\ \ref{r2_d}. We see in Fig.\ \ref{r2_d} that the constraints from the NIR flux agree with those from the rotational broadening (assuming the jet-binary alignment), but they impose only minor additional constraints. The current constraints could be improved with future more accurate measurements of the rotational broadening and the donor star's flux in the K band.

\section{The models of V1487 A\lowercase{ql}}
\label{models}

\subsection{The core mass--radius plane}
\label{core_radius}

\begin{figure}
\centerline{\includegraphics[width=\columnwidth]{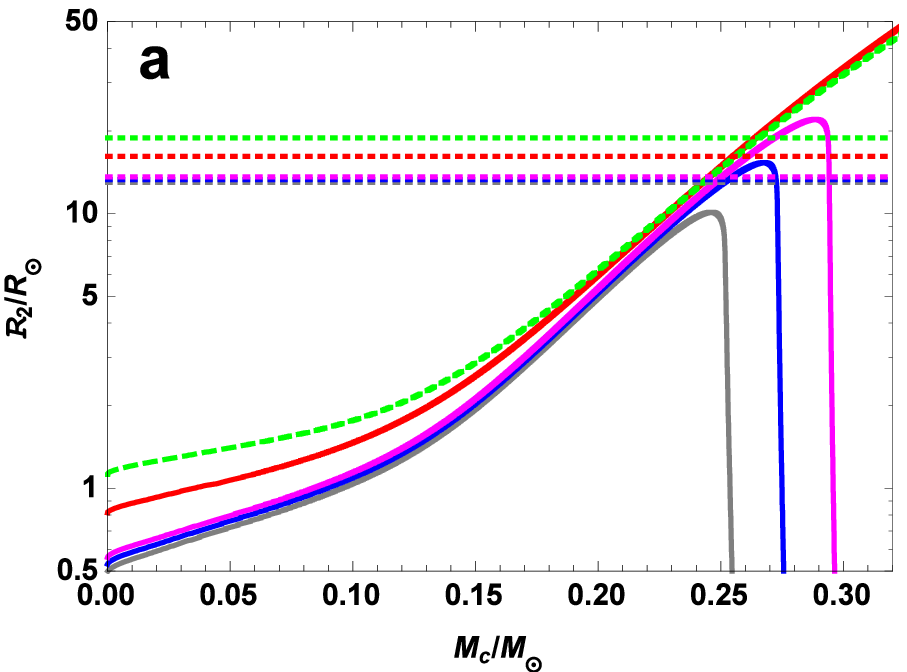}} 
\vskip 0.2cm
\centerline{\includegraphics[width=\columnwidth]{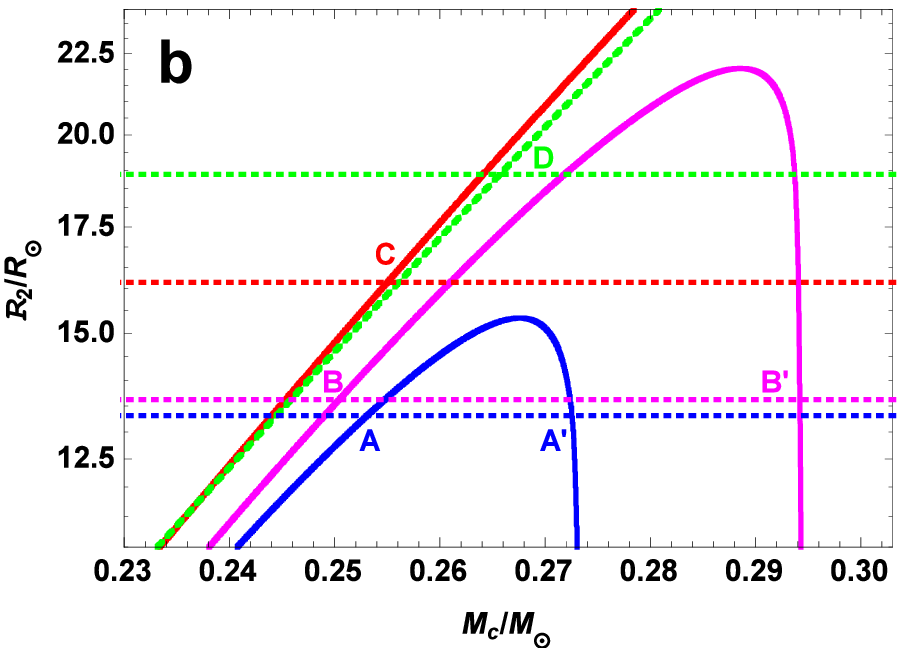}} 
      \caption{Evolution of partially stripped giants in the $M_{\rm c}$--$R_2$ diagram for $M_2= 0.26$, 0.28, 0.3 0.5 and $0.8 \msun$ for the grey, blue, magenta, red and green curves, respectively. The evolution proceeds (from left to right) at the constant total mass, during which the H-burning shell is moving outwards. This increases the mass of the He core and decreases the mass of the H-rich envelope. The horizontal lines (in respective colours) show the Roche-lobe radii for (from the top) $M_2 = 0.8$, 0.5, 0.3, 0.28 and $0.26 \msun$. Possible solutions are given by the intersections between a horizontal line and the corresponding evolutionary track. (a) The full studied parameter space. (b) A magnification of the intersection region. The possible solutions are marked with the letters: A and A$'$ for $M_2=0.28 \msun$, B and B$'$ for $0.3 \msun$, and C and D for 0.5 and $0.8 \msun$, respectively. The solutions A$'$ and B$'$ are unphysical as they do not assure a continuous mass transfer.}
     \label{rad_mc}
    \end{figure}

In order to calculate evolutionary models of stripped giants, we used the Warsaw stellar-evolution code (described in \citealt{z05}). The code was calibrated to reproduce the Sun at the solar age. This calibration resulted in the chemical composition of the H mass fraction of $X = 0.74$, the metallicity of $Z = 0.014$, and the mixing length parameter of $\alpha = 1.55$.

To reproduce the present state of the donor, we followed the evolution of a $1\msun$ star, which was maintained at a constant mass until hydrogen was nearly exhausted in its centre. Then, the mass removal from the surface started and continued until the donor star reached six different values of the mass, 0.26, 0.28, 0.3, 0.5, 0.8 and $0.9 \msun$. Until then, the He core had not formed. The further evolution was followed at a constant total mass. The H-burning shell moves outwards, increasing the mass of the He core, $M_{\rm c}$, and decreasing that of the H-rich envelope. Generally, this causes an increase of the radius of the stripped giant. However, when the mass of the remaining envelope gets sufficiently low, the giant starts to shrink, as shown in Fig.\ \ref{rad_mc}.

The particular choice of $1 \msun$ star for the starting configuration was dictated by the reason of the computational convenience.  Our motivation was to obtain thermal-equilibrium stars of a given mass. Since the structure of such stars does not depend on their evolutionary history, the choice of the initial configuration or the prescription for mass removal from the surface was unimportant\footnote{\citet{fm15} suggested that this initial configuration could involve a donor star as massive as $5 \msun$. The large amount of the mass from the donor accreted by black hole would then help to explain its claimed large spin. However, \citet{fm15} assumed fully conservative mass transfer throughout the system history, which included epochs of highly super-Eddington accretion, during which strong outflows most likely took place, questioning that assumption.}. To check that this is indeed the case, we made several experiments. First, during earlier calculations (described in \citealt{zdz16}) we calculated tracks for a $0.2 \msun$ stripped giant in two cases: one in which the mass removal from the initial $1\msun$ star started when it was near the end of the core H burning (but H was not yet completely exhausted), and second in which the mass removal started when  the initial $1 \msun$ star developed already a substantial ($0.12 \msun$) He core. We found that the structure and the further evolution of the $0.2 \msun$ stars was identical in both cases (for the core mass range of 0.12--$0.19\msun$). Second, we applied two different (arbitrary) rates of the mass removal, $1.3 \times 10^{-9}$ and $1.3 \times 10^{-8}\msun$/y, to the initial $1\msun$ star. Again, the resulting stripped giants were insensitive to these details. Finally, we followed the evolution of $0.5\msun$ stripped giant in two cases: one in which the starting configuration was $1\msun$ star and another in which it was a $1.4 \msun$ star. Again, both resulting tracks were identical. Additionally, we followed the evolution of 1.0 and $1.4 \msun$ stars without any mass removal (un-stripped giants).

The results of our calculations are presented in Fig.\ \ref{rad_mc}, which show the evolutionary tracks in the core mass--radius diagram. The tracks are shown for the stripped giants with the five lowest considered masses. The stars evolve at constant mass and the driving mechanism is the progress of the H-burning shell moving outwards. The radii of the partially stripped giants generally increase with $M_{\rm c}$, except for the shrinking when the masses of their envelopes become very low.

\begin{table}
\label{solutions}
\begin{center}
\caption{The parameters of the physical evolutionary models. The mass loss rate is in the unit of $10^{-9}\msun$/y.
}
\begin{tabular}{ccccccc}
\hline
Model & $M_2/\msun$ & $R_2/\rsun$ & $T_{\rm eff}$ [K] & $L_2/\lsun$ & $M_{\rm c}/\msun$ & $-\dot M_2$\\
\hline
A & 0.28 & 13.3 & 4160	& 47.1	& 0.2530  & 0.85\\
B & 0.30 & 13.6 & 4080	& 45.0	& 0.2504  & 1.56\\
C & 0.50 & 16.1 & 3970	& 57.3	& 0.2550  & 5.33\\
D & 0.80 & 18.9 & 4030	& 83.3	& 0.2656  & 13.0\\
E & 0.90 & 19.6 & 4050  & 92.0  & 0.2703  & 16.0\\
F & 1.00 & 20.3 & 4070  &100.8  & 0.2720  & 19.5\\
G & 1.40 & 22.8 & 4150  &135.9  & 0.2834  & 37.7\\
\hline
\end{tabular}
\end{center}
\end{table}

Fig.\ \ref{rad_mc} also shows values of the Roche-lobe radius of the donor calculated with the formula (\ref{r_k}) for the considered masses. Possible solutions that we consider as models for the donor are given by the intersections between a horizontal line (for a given mass) and the corresponding evolutionary track.

From Fig.\ \ref{rad_mc}, we immediately see that a stripped giant of the mass $0.26 \msun$ cannot provide a solution since during its evolution it never attains a sufficiently large radius. We have checked that the same is true for the mass of $0.27 \msun$. So, we are left with the remnants of the mass $\geq\!0.28 \msun$, which minimum value and its corresponding radius we show in Figs.\ \ref{r2_i} and \ref{r2_d}. In further discussion, we shall consider remnants of the masses $\gtrsim 0.28 \msun$. Magnified portions of the relevant tracks and horizontal lines from Fig.\ \ref{rad_mc}(a) are shown in Fig.\ \ref{rad_mc}(b). The possible solutions given by the intersections between the horizontal line (for a given mass) and the evolutionary track (for the same mass) are marked with consecutive capital letters, A, B, C and D for $M_2=0.28$, 0.3, 0.5 and $0.8\msun$, respectively, while the primed letters indicate the intersections during the final evolution stages. 

We point out that the solutions A$'$ and B$'$ are unphysical as they do not assure a continuous mass transfer between the components of the binary system since solutions lie on the declining parts of the evolutionary tracks. During this evolutionary phase, the star shrinks with the growing mass of the core. Therefore, any mass outflow would be quickly stopped. The parameters of the physical solutions are given in Table 1, including their effective temperature and luminosity. 

We note that the dependence of the stellar radius on the core mass can be considered for three different situations. First, we can consider the evolution of isolated giants at constant mass, as in Fig.\ \ref{rad_mc}. While this evolution does depend on the stellar mass, this dependence is relatively weak, and we can provide a fitting formula for the main dependence of $R_2(M_{\rm c})$ averaged over $M_2$, as also done by WRS83. Then, we can consider $R_2(M_{\rm c})$ during the mass transfer via Roche-lobe overflow, i.e., with a simultaneous decrease of the donor mass. This gives usually steeper dependencies. It corresponds to considering also the second term in equation (1) for $\dot R_2$ of WRS83. Finally, we can consider $R_2(M_{\rm c})$ for our solutions, imposing the stellar radius equal to that of the Roche lobe of a given mass. This does not correspond to any evolutionary sequence, but just parameterises our results for $P=33.85$ d. This, in turn, gives a flatter dependence. 

We compare our radius vs.\ the core mass dependencies with the results of WRS83 (and of \citealt{king93}, who retained only the first order in their formulae). They correspond to the first case above, i.e., for evolution of an isolated star, and averaging over the dependence on $M_2$. Given that we use different chemical composition and more contemporary physics (especially opacities), our results, while qualitatively similar, are quantitatively different. While WRS83 obtained $R_2\propto M_{\rm c}^{5.1}$, we find on average for $M_2\geq 0.5\msun$,
\begin{equation}
R_2\simeq 14\rsun [M_{\rm c}/(0.25\msun)]^{4.3},\label{r_mcore}
\end{equation}
for the evolution of a giant at a constant mass. We do see some dependence on $M_2$ in Fig.\ \ref{rad_mc}, especially for low masses (see also \citealt{zdz16}). Also, this dependence excludes ranges of $M_{\rm c}$ close to $M_2$, when the star starts to shrink. The second case above is discussed in Section \ref{outflow}. If we consider only the radii equal to the Roche-lobe radii at the period of \source\ (the third case), we find $R_2\propto M_{\rm c}^{3.2}$.

\subsection{The radius--luminosity plane}
\label{rad_lum}

So far, we have made use only of the value of the radius of the donor as a function of its mass. We can also use a second parameter, its luminosity, $L_2$. We present our treatment below, which currently leads to relatively loose constraints, given that $L_2$ has been so far only roughly estimated. However, a future more accurate measurement will lead to a precise determination of the mass.

\begin{figure}
\centerline{\includegraphics[width=\columnwidth]{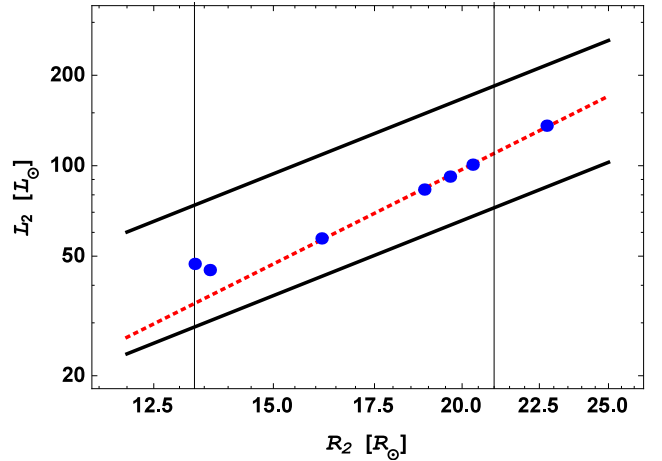}} 
\caption{A comparison of our solutions with the observational constraints in the $R_2$--$L_2$ plane. The heavy solid lines show the uncertainty range of $L_2$ due to the uncertainty of $T_{\rm eff}$, and the vertical thin lines show the minimum radius of $13.3\rsun$ (from our evolutionary calculations) and the maximum radius found to be allowed in Section \ref{radius}. The blue points marked show the positions of the solutions A--G (in order of increasing $R_2$). The red dotted line shows the fit to the cases with $M_2>0.3\msun$ (solutions C--G). }
\label{R_L}
\end{figure}

The luminosity can be estimated from the range of the allowed effective temperatures, $L_2=4\upi R_2^2 \sigma T_{\rm eff}^4$, where $\sigma$ is the Stefan-Boltzmann constant, and we use the values corresponding to giants of the spectral classes K0 III--M0 III, $T_{\rm eff}=3690$--4660 K \citep{cox00}. This uncertainty leads to the corresponding uncertainty of the luminosity estimate. The range of $L_2$ at a given $R_2$ corresponding to the adopted range of $T_{\rm eff}$ is shown in Fig.\ \ref{R_L}.

Our evolutionary model predicts the luminosity at any stage, and we compare the values corresponding to our solutions, see Table 1, with the above constraints in Fig.\ \ref{R_L}. We have found that all our solutions lie in middle of the allowed range, indicating a good agreement of our evolutionary calculations with the standard parameters of giants of \citet{cox00}, in spite of the reduction of the mass due to accretion mass loss\footnote{We do not show here the solution A$'$, which also agrees with above luminosity constraints. It was this solution that \citet{z15} advocated as the best model. Unfortunately, it was not noted at that time that this model did not assure a continuous mass transfer since it corresponded to the radius decreasing with time, and hence was unphysical.}. All of the solutions have the temperatures of $\simeq 3990$--4160 K, which correspond to the spectral class of K5 III \citep{cox00}. We thus find that the evolutionary considerations do not provide any significant upper limit on the mass and radius. For example, we have also calculated models with the final mass of $0.9\msun$ (solution E), $1.0\msun$ (solution F), and $1.4\msun$ (solution G), with the last one being well above the mass range allowed by the observations, and found that those solutions also lie in the middle of the range of $L_2$ allowed by the adopted range of $T_{\rm eff}$. We show their parameters in Table 1.

One comment that should be made about our models concerns the luminosity of a star losing mass through Roche lobe overflow. If the mass outflow is very rapid, the surface luminosity of the star may become significantly lower than that without outflow (as noted by \citealt*{greiner01a}). It is known that the black hole in the system is accreting matter roughly at the Eddington rate, $\sim 10^{-7}\msun$/yr, though the rate of the mass loss from the donor is much lower, see Section \ref{outflow}. However, we have found that even the Eddington rate is not high enough to decrease substantially the stellar surface luminosity. We superimposed a mass outflow at about Eddington rate on our model A, and found that the internal radiation flux in the outer layers started to decrease toward the surface, indicating that these layers were departing from thermal equilibrium. However, this departure was very small. The surface luminosity of the model with the outflow was smaller by less than 1 per cent compared with the model without outflow.

The obtained values of the luminosity show a similar steep dependence (excluding $M_2=0.28$ and $0.3\msun$) on the core mass to that of WRS83 of $L_2\propto M_{\rm c}^{8.1}$, but our normalization is $\sim$50 per cent higher,
\begin{equation}
L_2\simeq 50\lsun [M_{\rm c}/(0.25\msun)]^{8.15},
\label{l_mcore}
\end{equation}
with the relative error $<$3 per cent. Combining it with the dependence of $R_2$ assuming the Roche-lobe radius for the observed $P$ (the third case at the end of Section \ref{core_radius}), we find $L_2\propto R_2^{2.5}$, which is shown in Fig.\ \ref{R_L}.

\section{The mass outflow rate from the donor}
\label{outflow}

\begin{figure}
\centerline{\includegraphics[width=\columnwidth]{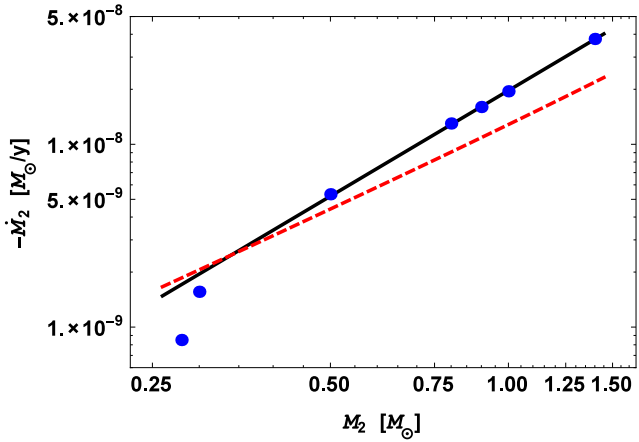}} 
\caption{The mass transfer rate vs.\ the stellar mass. The blue dots show the rates obtained in our calculations, and the solid line shows the dependence fitted to the cases with $M_2\geq 0.5\msun$, equation (\ref{mdot_w}). The dashed line shows the dependence of equation (25a) of WRS83. The mass range allowed for \source\ is $0.28\lesssim M_2/\msun\lesssim 1$.
}
\label{mdot_m2}
\end{figure}

Having constructed the models describing the internal structure of  the donor, we can calculate the rate of the mass transfer between the components of the binary system implied by our models. To do so, let us locate each of our models in a binary with $P=33.85$ d and a $12.4 \msun$ black hole. We assume the conservative mode of the mass transfer (conservation of the total mass and of the total orbital angular momentum). Then, we calculate numerically at which rate of the mass outflow from the star the changes of the stellar radius will follow the changes of the Roche lobe around it. The resulting rates are given in Table 1 and shown in Fig.\ \ref{mdot_m2}, where we see a strong dependence of $\dot M_2$ on $M_2$. 

In addition to the Roche-lobe overflow, the donor loses mass via stellar wind, whose rate can be estimated using equation (106) of \citet*{hpt00}. This gives $-\dot M_{\rm 2,wind}\simeq (3.7$--$4.5)\times 10^{-10}\msun\,{\rm y}^{-1}$ for our considered cases. We see it is much less than the Roche-lobe rate in most cases, and comparable to it only for the lowest allowed mass. Given that only a small fraction of the wind is captured by the black hole, we neglect its effect hereafter.

Our numerical results can be compared with the theoretical expression for $\dot M_2$ separated into three physical terms,
\begin{equation}
\dot M_2 =\left(\frac{{\rm d}R_2}{{\rm d}M_2}\right)^{-1} \frac{{\rm d}R_2}{{\rm d}M_{\rm c}} \dot M_{\rm c},
\label{mdot}
\end{equation}
where ${\rm d}R_2/{\rm d}M_2$ is the rate of the Roche-lobe changes with the changing mass, given by equation (12) of WRS83, with $R_2\simprop M_2^{-5/3}$ at $M_2\ll M_1$, and ${\rm d}R_2/{\rm d}M_{\rm c}$ is the derivative along the solution with the mass transfer, i.e., the second case discussed in Section \ref{core_radius}. We find $R_2\propto M_{\rm c}^{c_1}$ along the solutions with the mass transfer, and $c_1\simeq 4.53$--4.88 for $M_2=0.5$--$1.4\msun$. The time derivative of the core mass is linked to the luminosity and the efficiency of hydrogen burning (accounting for neutrino losses), $\epsilon_{\rm H}\simeq 6.0\times 10^{18}$ erg/g (as given by WRS83), $-\dot M_{\rm c}= L/(X' \epsilon_{\rm H})$. Here, $X'$ is the H content immediately above the H-burning shell, which, due to the evolutionary H burning, can be substantially lower than the intial $X=0.74$ (which still approximately equal that at the stellar surface). In our cases with $M_2=0.5$--$1.4\msun$, $X'\simeq 0.62$--0.63. We find the values of $\dot M_2$ obtained from equation (\ref{mdot}) using the obtained values of $c_1$ and $X'$ to agree very well with our numerical results, shown in Table 1 and Fig.\ \ref{mdot_m2}. We note that $\dot M_2\rightarrow 0$ at the exact minimum possible mass, because then ${\rm d}R_2/{\rm d}M_{\rm c}\rightarrow 0$. Thus, arbitrarily low values of $\dot M_2$ can be obtained when the minimum allowed mass is approached, but this requires significant fine-tuning, see Fig.\ \ref{rad_mc}. We thus see that for the allowed range of $M_2$ of 0.28--$1\msun$, $-\dot M_2\lesssim 2\times 10^{-8}\msun\,{\rm y}^{-1}$. 

We can compare our values with fitting expressions given in literature. The rate given by equation (15) of WRS83 in terms of $M_2$ and $M_{\rm c}$ (and with a very weak dependence on $M_1$) is equivalent to our equation (\ref{mdot}), but we have now different numerical coefficients appearing in that formula. The formula (25a) in WRS83, giving $\dot M_2$ in terms of $M_2$ and $P$ (with the same weak dependence on $M_1$ as in their eq.\ 15) is shown by the dashed line in Fig.\ \ref{mdot_m2}. We see it has a different slope from that shown by our values. We have fitted the values of $\dot M_2$ for $M_2\geq 0.5\msun$, and obtained
\begin{equation}
-\dot M_2\simeq 7.4\times 10^{-10} \left(\frac{P}{1\,{\rm d}}\right)^{0.93} \left(\frac{M_2}{\msun}\right)^{1.9} \msun\,{\rm y}^{-1},
\label{mdot_w}
\end{equation}
which is shown in Fig.\ \ref{mdot_m2} by the solid line. The dependence on $P$ has been kept unchanged with respect to WRS83. The disagreement at the lowest masses is a boundary effect, caused by $\dot M_2\rightarrow 0$ at the exact minimum possible mass (as discussed above). Our results supersede those of \citet{vilhu02}, who used the fits of WRS83 with a previous estimate of $M_2\simeq (1.2\pm 0.2)\msun$.

We compare the obtained mass transfer rates with the average accretion rate required to power the outburst of \source. We used for that the results of \citet*{dwg04}. From their fig.\ 5, we infer the average observed bolometric flux corresponding to $\sim$3/4 of the Eddington luminosity for $M_1=14\msun$, pure H and at $d=12.5$ kpc assumed by them, which is $7.1\times 10^{-8}$ erg cm$^{-2}$ s$^{-1}$. This corresponds to $L_1\simeq 6.2(d/8.6\,{\rm kpc})^2\times 10^{38}$ erg s$^{-1}$, while \citet{vilhu02} estimated $L_1\simeq 5.7(d/8.6\,{\rm kpc})^2\times 10^{38}$ erg s$^{-1}$, a very similar value. We thus adopt $L_1= 6(d/8.6\,{\rm kpc})^2\times 10^{38}$ erg s$^{-1}$, which corresponds to the accretion rate of $1.1(d/8.6\,{\rm kpc})^2 (\epsilon/0.1)^{-1} \times 10^{-7}\msun/$y, where $\epsilon$ is the accretion efficiency. The accreted mass during the 25 y of the outburst is $\simeq 2.7(d/8.6\,{\rm kpc})^2 (\epsilon/0.1)^{-1} \times 10^{-6}\msun$. See \citet{truss06} for estimates of the disc mass in \source.

At the observational best-fit donor mass value of $M_2\simeq 0.5\msun$ and at our corresponding value of $\dot M_2$ (Table 1), the duty cycle (defined as the fraction of the time spent in outbursts) is thus about 5 per cent. At the maximum allowed mass of $1\msun$, it is 20 per cent, while it is much less at the lowest allowed donor masses. 

Our estimated range of the mass transfer rate is much below the critical rate at which the system would become persistent. \citet*{coriat12} have obtained the critical rate as $\simeq 4\times 10^{15} (P/1\,{\rm h})^{1.59}\,{\rm g\,s}^{-1}$, which gives $-\dot M_{\rm crit}\simeq 3\times 10^{-6} \msun\,{\rm y}^{-1}$ for \source, i.e., much more than our estimated rates (and much more than even the accretion rate during the outburst). The transientness parameter, defined by \citet{coriat12} as $\dot M_2/\dot M_{\rm crit}$, is thus $< 7\times 10^{-3}$. The comparison with other black-hole transients shown in their fig.\ 4 suggests then the average recurrence time of at least 30 y, with a large uncertainty, and consistent with that estimated from the duty cycle of $\gtrsim$130 y.

A potentially important effect we have neglected is the illumination of the donor by the X-ray source (pointed out by \citealt{vilhu02}). At the radius for $M_2=0.5\msun$, the donor subtends a solid angle of $0.006\times 4\upi$ (see equation \ref{kepler}), and the ratio of the illuminating luminosity (neglecting shielding and assuming isotropy) to $L_2/2$ (emitted by the hemisphere facing the black hole) is 33. At the estimated duty cycle, the average illuminating luminosity is larger than the intrinsic one, and illumination will slightly increase the stellar radius, which will in turn increase the mass transfer rate (e.g., \citealt{gk93}). However, a detailed treatment of donor illumination, including calculations of the albedo and the time dependence, is beyond the scope of this paper. We note only that such an effect appears to be clearly seen in another black-hole binary with a giant donor, GX 339--4, which has $P\simeq 1.76$\,d \citep{hynes03} and $M_2\lesssim 1\msun$ \citep*{munoz08}. Given its frequent outbursts, its average mass transfer rate required to power the outbursts can be relatively accurately estimated as $\sim 10^{-8} \msun\,{\rm y}^{-1}$ \citep{zdz04,coriat12}. On the other hand, the theoretical mass transfer rate based on the formalism of WRS83 is $\lesssim 10^{-9} \msun\,{\rm y}^{-1}$ \citep{munoz08}, see also equation (\ref{mdot_w}), which discrepancy is likely to be accounted for by illumination.

\section{Conclusions}

We have studied constraints on the present mass and radius of the donor in \source/V1487 Aql by constructing evolutionary models of the internal structure of this star. We have compared the resulting radii and luminosity of our physical models to observational constraints. We find solutions with stripped giants of the mass of $\geq\!0.28\msun$ and of the spectral class K5 III, independent of the distance to the system, and a distance-dependent upper limit, $\lesssim\!1\msun$. We have also obtained the constraints from the observed NIR magnitude of the companion, which are found to be consistent with the other constraints. Overall, we have found very good agreement of our theoretical models with observational constraints.

Then, we have estimated the mass transfer rate, which we found to be $\lesssim 2\times 10^{-8}\msun\,{\rm y}^{-1}$. Comparing to the average luminosity of \source, we obtain the duty cycle of $\lesssim 20$ per cent. Thus, the expected average recurrence time is $\gtrsim 130$\,y. The mass transfer rate and duty cycle become four times lower at $M_2\simeq 0.5\msun$, which is the observationally-determined best-fit mass. The obtained allowed values of the mass transfer rate are more than two orders of magnitude below the critical rate, at which the system would become persistent. 

The main quantities that would profit from more accurate measurements are the rotational broadening, which would allow a better constraint on the mass ratio, and the donor luminosity, which would allow a direct comparison with the predictions of our theoretical model. Both quantities can be more precisely measured after \source\ transits to quiescence. However, the time of this future event remains currently uncertain \citep{truss06}.

\section*{Acknowledgements}

We thank Joanna Miko{\l}ajewska and Tomaso Belloni for valuable discussions, and the referee for valuable suggestions. This research has been supported in part by the Polish National Science Centre grants 2013/10/M/ST9/00729 and 2015/18/A/ST9/00746.

\label{lastpage}
\end{document}